\documentclass[a4paper, 10pt, aps, nofootinbib, twocolumn]{revtex4-1}

\usepackage{graphicx}
\usepackage{amsfonts}
\usepackage{amssymb}
\usepackage{amsbsy}
\usepackage{amsmath}
\usepackage{mathrsfs}
\usepackage{latexsym}
\usepackage{bm}
\usepackage{subfigure} 
\usepackage{wasysym}
\usepackage{mathbbol}
\usepackage{bigints} 
\usepackage{fontawesome}
\usepackage{natbib}
\usepackage{xcolor}
\usepackage{footnote}
\usepackage[colorlinks = true, linkcolor=blue, citecolor=cyan, urlcolor=blue]{hyperref}
\usepackage{orcidlink}

\begin{document}
\title{Transonic accretion flow in the mini discs of a binary black hole system}

\author{Subhankar Patra \orcidlink{0000-0001-7603-3923}}
\email{psubhankar@iitg.ac.in}

\author{Bibhas Ranjan Majhi \orcidlink{0000-0001-8621-1324}}
\email{bibhas.majhi@iitg.ac.in}

\author{Santabrata Das \orcidlink{0000-0003-4399-5047}}
\email{sbdas@iitg.ac.in}

\affiliation{Department of Physics, Indian Institute of Technology Guwahati, Guwahati 781039, Assam, India}

\date{\today}

\begin{abstract}
	We study the general relativistic transonic accretion flow around the primary black hole, which forms the circumprimary disc (CPD), within a binary black hole (BBH) system. The BBH spacetime is characterized by the mass ratio ($q$) and the separation distance ($z_2$) between the two black holes. We numerically solve the radial momentum and energy equations to obtain the accretion solutions. It is observed that the CPD can exhibit shock solutions, which exist for a wide range parameter space spanned by flow specific angular momentum ($\lambda$) and energy ($E$). We find that the shock parameter space is modified by $q$ and $z_2$. Investigations show that $q$ and $z_2$ also affect various shock properties, such as density compression and temperature compression across the shock fronts. Moreover, we calculate the spectral energy distributions (SEDs) of the CPD and examine how the SEDs are modified by $q$ and $z_2$ for both shock-free and shock-induced accretion solutions. SED is found to be nearly independent of the binary parameters. We essentially show that although $q$ and $z_2$ alter the effective horizon area of the primary black hole located at the center of the CPD, they have a minimal impact on the dynamical and spectral properties of the accretion flow around the primary black hole.   
\end{abstract}





\maketitle
\section{Introduction}
\label{sec:introduction}

Binary black hole (BBH) systems are powerful multi-messenger astrophysical sources of gravitational waves (GWs) and electromagnetic (EM) waves. The formation scenarios of BBHs are not clearly understood yet. In this regard, various astrophysical channels have been proposed in the literature. Notable studies by \cite{Begelman-1980-307} and \cite{Roos-1981-218} indicate that supermassive black hole binaries (SMBHBs) form as a result of galaxy mergers. On the other hand, stellar-mass BBHs are generated through dynamic encounters in the environments of dense star clusters \cite[]{Banerjee-2017-524, Rodriguez-2018-151101}. Recently, it has been demonstrated that the gaseous dynamical friction can form a BBH system in the active galactic nucleus (AGN) discs \cite[]{DeLaurentiis-2023-1126}.

In addition to the understanding of the formation channels of BBHs, detecting them through observations is also a vital aspect in astrophysics. Meanwhile, LIGO and Virgo collaborations reveal that the gravitational interactions within BBHs can produce GWs \cite[]{Abbott-2016-061102}, as predicted by general relativity, thereby providing direct evidence of stellar-mass BBHs through GWs.  On the other hand, SMBHBs at sub-parsec separations are the strong GW sources \cite[]{Burke-2018-5}, which may be detectable by the Pulsar Timing Arrays \cite[]{Arzoumanian-2020-L34} or the future mission Laser Interferometer Space Antenna \cite[]{Amaro-2017-00786, Mangiagli-2022-103017}. Like AGNs, BBHs can radiate across the electromagnetic spectrum due to the accretion of matter from the surroundings. Therefore, in addition to GWs, analyzing EM signature from BBHs is also crucial for their study. In recent years, several EM observational works, including both the thermal and non-thermal emissions, have come out for the BBHs \cite[]{Roedig-2014-115, Orazio-2015-2540, Graham-2015-1562, Graham-2015-74, Liu-2019-36, Saade-2020-148, Bogdanovic-2021-3, Gutierrez-2022-137, Lee-2024-141, Mondal-2024-A279}. Indeed, these investigations provide the valuable information about the binary evolution, mass accretion rate, disc inclination angle, mass and the spin of the individual black holes, etc.

The accretion flow onto a BBH system is qualitatively different from a single black hole accretion. For BBH accretion, a truncated disc surrounds the entire binary system \cite[]{Gutierrez-2024-14843}, called the circumbinary disc (CBD). The gravitational potential of the binary determines the orbital structure of the CBD, while viscous stress enables accretion by transporting angular momentum outward \cite{Miranda-2017-1170}. However, in the presence of a binary, the disc is truncated at a certain radius, which defines its inner boundary, where the outward tidal torques exerted by the binary balance the inward viscous torques \cite{Papaloizou-1977-441, Lin-1986-846, Artymowicz-1994-651, Miranda-2017-1170, Heath-2020-A64}. This balance typically occurs at Lindblad resonances \cite{Artymowicz-1994-651}, where disc material resonates with the binary's orbital motion, resulting in the formation of a low-density cavity around the binary. The amount of matter that falls into the cavity depends on the disc scale height, thermodynamical properties (e.g., temperature and pressure gradients), and flow properties (e.g., viscosity) \cite{Artymowicz-1996-L77, D-Orazio-2013-2997, Heath-2020-A64, Tiede-2020-43, Tiede-2022-24, Tiede-2025-144}. The matter can flow into the cavity through narrow streams and begin to accrete onto the individual black holes, leading to the formation of a pair of mini-discs. Various recent studies have shown that this matter inflow can be highly variable and may exhibit characteristic periodicity depending on binary mass ratio and binary orbital parameters (e.g., eccentricity, orbital period) \cite{D-Orazio-2013-2997, Tiede-2025-144}. The mini discs surrounding the primary (more massive) and secondary (less massive) black holes are referred to as circumprimary discs (CPDs) and circumsecondary discs (CSDs), respectively. The presence of multiple discs (i.e., CBD, CPD, and CSD) complicates the study of BBH accretion compared to the accretion of individual black holes. Numerous theoretical models on BBHs accretion flow have been proposed in Newtonian gravity and general relativity (GR) \cite[and references therein]{Sesana-2011-L35, Bowen-2019-76, Armengol-2021-16, Combi-2022-187, Gutierrez-2022-137, Lee-2024-141,Gutierrez-2024-506, Gutierrez-2024-14843}. All these models based on different physical conditions and explained the dynamics of CBD, mini discs, role of magnetic field, outflows or jets and EM spectral signature, etc., which indeed help to distinguish them from the ordinary AGNs. However, to the best of our knowledge, no body has conveyed the transonic acccretion flow for the BBH systems, where flow must satisfied the inner boundary condition set by the horizon \cite[]{Liang-1980-271, Abramowicz-1981-314}. Toward this, for the first time we study the GR hydrodynamics of transonic accretion flow in the background of a BBH metric \cite[]{Astorino-2021-136506}, which represents a system of two Schwarzschild black holes in an external gravitational field. We expect this study may provide the better understanding of the BBHs.

Finding an exact analytical solution for binary or multiple black hole sources in GR is very challenging. Several attempts have been made to address this problem \cite[]{Kramer-1980-259, Dietz-1985-319}, where it often requires the extra matter fields, commonly interpreted as struts or cosmological strings, to support the gravitational attraction between the black holes. However, all of them are hard to believe as a proper manifold because some fundamental theoretical issues arise regarding asymptotic flatness, singular behaviors, validity of energy conditions, etc., along with the stability issues. In \cite{Majumdar-1947-390, papapetrou-1945-191, Hartle-1972-87, Emparan-2000-104009}, some regular metric solutions were proposed for the multi-black hole configurations. In these models, electromagnetic fields or charged matter help to maintain the equilibrium between the black holes. However, these systems are generally unstable to small perturbations, which could cause them to merge or drift apart. \citet{Astorino-2021-136506} were the first to propose an exact analytical solution for a configuration of two static black holes, where an external back-reacting gravitational field maintains the equilibrium between two sources without requiring additional matter fields. Most importantly, this metric is regular everywhere outside the event horizons, i.e., free from any conical and curvature singularities, providing a physical BBH metric solution. In their construction, they employed the Ernst framework, a formalism for generating exact stationary and axisymmetric solutions by reducing the Einstein field equations to a complex scalar equation for the Ernst potential. Within this framework, they used a technique in which the external gravitational field is modeled as a series of multipole moments. By appropriately choosing the multipole structure, they were able to achieve mutual equilibrium between the black holes against their gravitational attraction and also avoid any curvature singularities outside the horizons. In this way, they constructed a physically acceptable exact solution representing a realistic binary black hole configuration. It is characterized by the masses and positions of the two black holes, external field parameters, and a gauge parameter. In this work, we investigate the transonic accretion flow around the primary black hole in the background of this metric. We find both the shock-free and shock-induced accretion solutions, and examine their corresponding disc properties, such as density distribution, temperature distribution, and emissivity, etc. Since the horizon of individual Schwarzschild black holes are distorted by their mutual interactions, the presence of the secondary black hole may influence the accretion properties of the CPD. To address this possibility, we explore the impact of binary parameters, specifically the mass ratio ($q$) of the two black holes and their separation ($z_2$) on the accretion solutions. We also examine the various shock properties and luminosity distributions of the CPD as functions of $q$ and $z_2$. These investigations show that the accretion properties of the CPD are nearly independent of the binary parameters, featuring earlier predictions based on different aspects of binary accretion \cite[]{Lee-2024-141}.

The paper is organized as follows. In Section \ref{sec:BBH-spacetime}, we introduce the BBH spacetime. The model equations governing the accretion flow in the CPD are formulated in Section \ref{sec:model-equations}. In Section \ref{sec:results}, we present the transonic accretion solutions and discuss the impact of binary parameters on various accretion properties. Finally, we present the concluding remarks in Section \ref{sec:conclusions}.

\section{Binary black hole spacetime} 
\label{sec:BBH-spacetime} 
Astrophysical black holes are always embedded in external gravitational fields, which can be described by a series of multipole moments. It has been shown in \cite{Castro-2011-225020} that any non-spherical matter distributions, such as accretion discs, galaxies, nebulae, etc., will generally produce a multipolar gravitational field. These external gravitational fields provide a physical mechanism to maintain the equilibrium between the black holes. The line element of a system of two Schwarzschild black holes immersed in a dipole-quadrupole external gravitational field can be expressed in cylindrical Weyl coordinates ($t, x, \phi, z$) as \cite[]{Astorino-2021-136506},
\begin{equation}
		\begin{split}
			\label{eq:BBH-metric}
			ds^{2} & = -V(x, z)dt^{2} + x^{2}V^{-1}(x, z)d\phi^{2}\\
			& + f_0(x, z)(dx^{2} + dz^{2}),
		\end{split}
\end{equation}
where
\begin{equation}
	\begin{split}
		V (x, z) & = \frac{\mu_1\mu_3}{\mu_2\mu_4} \exp \biggl[2b_1z + 2b_2\biggl(z^2 - \frac{x^2}{2}\biggr)\biggr] , \\
		f_0(x, z) & = 16C_f \frac{\mu_1^3\mu_2^5\mu_3^3\mu_4^5}{W_{11}W_{22}W_{33}W_{44}W_{13}^2W_{24}^2Y_{12}Y_{14}Y_{23}Y_{34}}\\
		& \times \exp\biggl[-b_1^2x^2 + \frac{b_2^2}{2} \bigl(x^2 - 8z^2\bigr)x^2 - 4b_1b_2 z x^2 \\
		& + 2 b_1 (-z + \mu_1 - \mu_2 + \mu_3 - \mu_4 )\\
		& + b_2 \bigl(-2z^2 + x^2 + 4z (\mu_1 - \mu_2) + \mu_1^2 - \mu_2^2\\
		& + (\mu_3 - \mu_4) (4z + \mu_3 + \mu_4) \bigr)\biggr].
	\end{split}
\end{equation}
In the above expressions, $W_{ij} = x^2+\mu_i\mu_j$, $Y_{ij}=(\mu_i-\mu_j)^2$, and $\mu_i=\sqrt{x^2+(z-w_i)^2}-(z-w_i)$, where ($i, j$) take values from $1$ to $4$. The values of $w_i$ are chosen as $w_1 = z_1 - m_1$, $w_2 = z_1 + m_1$, $w_3 = z_2 - m_2$, and $w_4 = z_2 + m_2$.  Here, the quantities $m_i$ and $z_i$ denote the mass and position of the $i$-th black hole, respectively. Additionally, $b_1$ and $b_2$ are the dipole and quadrupole moments in multipole expansion of the external gravitational field, respectively, while $C_f$ refers the gauge parameter.

\begin{figure}
	\centering
	\includegraphics[width=0.8\columnwidth]{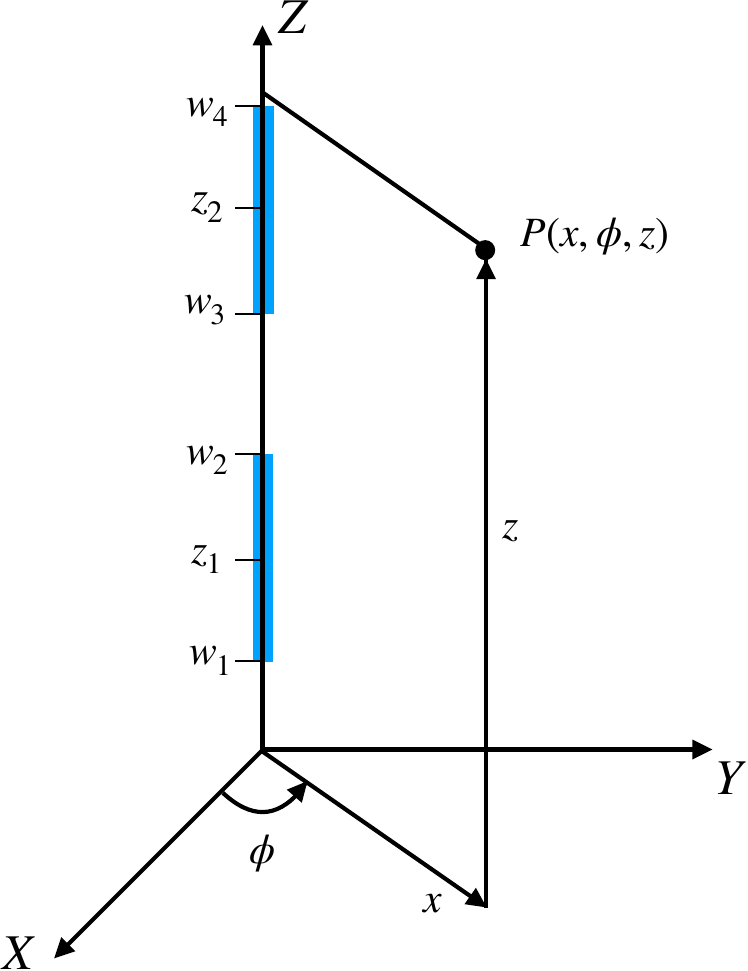}
	\caption{Schematic diagram of a binary black hole system featuring two co-linear Schwarzschild black holes. The primary and secondary black holes are located at $z_1$ and $z_2$, respectively. The thick blue lines represent the extended regions of the event horizons along the $Z$-axis. For the primary black hole, the horizon extends between $w_1$ and $w_2$, while for the secondary black hole, the horizon extends between $w_3$ and $w_4$. See the test for more details.}
	\label{fig:BBH}
\end{figure}

The static and axisymmetric metric (\ref{eq:BBH-metric}) is an exact analytical solution of the vacuum Einstein's equations in general relativity. It is important to note that the balance between two sources would be lost when the external gravitational field is absence ($b_1 = b_2 = 0$). In such a scenario, Eq.~(\ref{eq:BBH-metric}) boils down to the usual Bach-Weyl metric \cite[]{Bach-1922-134, Israel-1964-331, Costa-2000-469}, which also represents the double Schwarzschild black hole solution. In the binary system (\ref{eq:BBH-metric}), two event horizons extend into the regions $w_1 < z < w_2$ and $w_3 < z <w_4$ when $x = 0$, covering the curvature singularities of the spacetime. Moreover, it is not affected by any conical singularities along the $Z$-axis under the regularity condition $f_0V \rightarrow 1$ as $x \rightarrow 0$ \cite[]{Alekseev-2019-68}, which is valid everywhere outside the event horizons of individual sources. To satisfy the equation $f_0V = 1$, we consider a particular model of the external gravitational background where the parameters $C_f$, $b_1$, and $b_2$ are expressed in terms of intrinsic black hole parameters ($m_1$, $m_2$, $z_1$, $z_2$) as \cite[]{Astorino-2021-136506},
\begin{equation}
	\begin{split}
		C_f & = 256 m_1^2 m_2^2 (m_1+m_2+z_1-z_2)^2\\
		& \times (m_1+m_2-z_1+z_2)^2 ,\\
		b_1 & = -\frac{(m_1z_1+m_2z_2)}{4m_1m_2(z_1-z_2)}\\
		& \times \log\biggl[\frac{(m_1-m_2+z_1-z_2)(m_1-m_2-z_1+z_2)}{(m_1+m_2+z_1-z_2)(m_1+m_2-z_1+z_2)}\biggr] ,\\
		b_2 & = \frac{(m_1+m_2)}{8m_1m_2(z_1-z_2)}\\
		& \times \log\biggl[\frac{(m_1-m_2+z_1-z_2)(m_1-m_2-z_1+z_2)}{(m_1+m_2+z_1-z_2)(m_1+m_2-z_1+z_2)}\biggr] .
	\end{split}
\end{equation}
It is worth pointing out that $m_1$, $m_2$, $z_1$, and $z_2$ values remain unconstrained, providing a more generic double black hole configuration. In Fig.~\ref{fig:BBH}, we display a binary black hole system in coordinates ($t, x, \phi, z$). 

\begin{figure}
	\centering
	\includegraphics[width=\columnwidth]{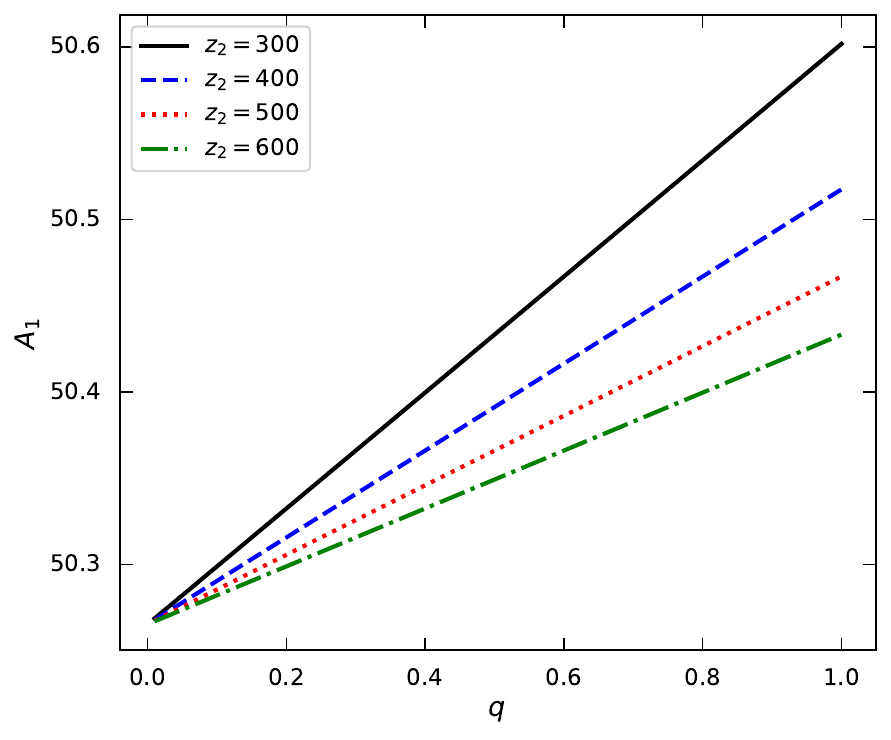}
	\caption{Plot of horizon area ($A_1$) of the primary black hole with binary mass ratio ($q$) for different binary separations $z_2 = 300$, $400$, $500$, and $600$. See the text for details.}
	\label{fig:A1}
\end{figure}

To analyze the event horizons of the two sources, we need to consider the near-horizon limit of metric (\ref{eq:BBH-metric}). The near-horizon geometry of the primary black hole has been examined in \cite{Astorino-2021-136506}. In that study, the coordinates transformation $ x = \sqrt{r(r-2m_1)}\sin\theta$ (event horizon at $x_{\rm H} = 0$) and $z = z_1+(r-m_1)\cos\theta$ brings the metric (\ref{eq:BBH-metric}) into a deformed Schwarzschild metric as $r$ approaches $2m_1$. The expression of this deformed Schwarzschild metric is given by \cite{Astorino-2021-136506},
	\begin{equation}
		\begin{split}
			ds^2 & \simeq h_0(\theta)\left[-\left(1 - \frac{2m_1}{r}\right)e^{F_1(\theta)} dt^2 + \frac{D^2e^{F_2(\theta)}}{1 - \frac{2m_1}{r}}dr^2\right]\\
			& + (2m_1)^2\left[h_0(\theta)D^2e^{F_2(\theta)}d\theta^2 + \frac{\sin^2\theta}{h_0(\theta)e^{F_1(\theta)}}d\phi^2\right],
		\end{split}
		\label{eq:deformed-Sch-metric}
	\end{equation}
	where
	\begin{equation}
		\begin{split}
			\label{eq:deformed-Sch-metric-1}
			h_0(\theta) & = \frac{m_1\cos \theta + m_2 + z_1 - z_2}{m_1\cos \theta - m_2 + z_1 - z_2}, \\
			D & = \frac{m_1 + m_2 - z_1 + z_2}{m_1 - m_2 - z_1 + z_2},\\
			F_1(\theta) & = 2\left[b_1 + b_2(z_1 + m_1\cos \theta)\right](z_1 + m_1\cos \theta),\\
			F_2(\theta) & = 2b_1 [m_1 \cos \theta - 2m_1 - 4m_2 - z_1]\\
			& + 2b_2[m_1^2 \cos^2\theta + 2m_1z_1\cos \theta \\ & - 2m_1^2 - z_1^2 - 4m_1z_1 - 8m_2z_2].
		\end{split}
	\end{equation}
	Note that when both the external field and the second black hole vanish, one finds from Eq. (\ref{eq:deformed-Sch-metric-1}) that $h_0(\theta) = 1$, $D = 1$, $F_1(\theta) = 0$, and $F_2(\theta) = 0$. Therefore, Eq. (\ref{eq:deformed-Sch-metric}) reduces to the usual Schwarzschild black hole of mass $m_1$. Also, from Eq. (\ref{eq:deformed-Sch-metric}) it is evident that both the external fields and the secondary black hole can distort the primary black hole horizon. A similar description applies to the secondary black hole as well. Indeed, the binary black hole system (\ref{eq:BBH-metric}) contains two distorted Schwarzschild black holes. The horizon area of primary black hole is calculated as \cite[]{Astorino-2021-136506},
\begin{equation}
	\begin{split}
		\mathcal{A}_1 = 16\pi \frac{1 + q + z_2}{1 - q + z_2} \exp [-2b_1 (1 + 2q) - 8b_2qz_2],
	\end{split}
\end{equation}
where $q$ ($= m_2/m_1$) is defined as the binary mass ratio. In this work, we use a unit system: $G = c = m_1 = 1$, where $G$ is the gravitational constant and $c$ is the speed of light. Also, we place the primary black hole at $z_1 = 0$. In Fig.~\ref{fig:A1}, we illustrate the dependency of $A_1$ on mass ratio ($q$) and distance between two sources ($z_2$), where the variation of $A_1$ as a function of $q$ for different $z_2$ values has been displayed. The solid (black), dashed (blue), dotted (red), and dash-dotted (green) curves represent the obtained results for $z_2 = 300$, $400$, $500$, and $600$, respectively. We observed that for a given $z_2$, $A_1$ increases with $q$. However, for a fixed $q$, $A_1$ decreases with $z_2$. Therefore, $q$ and $z_2$ play opposite roles in determining the characteristics of $A_1$.

In the introduction section, we provide a detailed discussion on how the authors in \cite{Astorino-2021-136506} proposed a static BBH metric by considering phenomenologically realistic scenarios as much as possible. In that study, to maintain the equilibrium between the black holes against the gravitational attraction between them, the authors did not include singular matter (cosmological strings) or charged matter as these matter fields routinely introduce the singularities outside the event horizons. Instead, they considered the black holes to be immersed in an external gravitational field. Specifically, they included only the dipole and quadrupole moments in the multipolar expansion of the external field, as these are sufficient to stabilize the BBH system. Also, the multipolar expansion is a key approach for regularizing the nodal singularity, due to it's foundation in Ernst's insight. However, as the masses of the black holes continuously increase due to the accretion of matter from the surroundings, it becomes challenging for the external fields to maintain stability in the system. When the black holes masses grow sufficiently, such that the gravitational attraction between them exceeds the repulsion from the external fields, the BBH system will collapse into a single black hole. However, the merging of black holes is an adiabatic process, i.e., it will take a very long time ($\sim 10^7 - 10^9$ yr) to occur, as the mass accretion rates of supermassive black holes are very small \cite[]{Yarza-2020-50}. Over this extended timescale, the system is expected to remain nearly stable, with the two black holes gradually drifting toward each other in a manner that is nearly negligible. Also, in our work, we focus on scenarios with large binary separations. Therefore, when studying the accretion physics over such a prolonged timescale, the metric we have used is expected to closely resemble the nearly static configuration of a BBH system. However, in a real BBH accretion scenario, applying the metric (\ref{eq:BBH-metric}) would be challenging, as the system is inherently dynamic. Since the primary goal of this work is to investigate how the presence of a secondary black hole influences the event horizon of the primary black hole and, consequently, its accretion properties, we adopt this simplified BBH metric.

\section{Model equations}
\label{sec:model-equations}
In this section, we aim to model the transonic accretion flow in the CPD within the stationary and axisymmetric BBH system (\ref{eq:BBH-metric}). Hydrodynamics of the CPD are studied within the framework of general relativity \cite[]{Rezzolla-2013-book}. We consider the fluid motion is restricted on the $z = 0$ plane of the central source ($u^z = 0$ and $\partial Q/\partial z = 0$, where $Q$ represents any flow parameter). It is important to note that the assumption of $u^z = 0$ is a valid approximation when the binary mass ratio ($q$) is small and the binary separation ($z_2$) is large. This is because the gravitational force exerted by the secondary black hole on the accreting material surrounding the primary black hole is negligible compared to the force from the primary black hole. However, for moderate to large mass ratios (up to a maximum value of $1$), assuming $u^z = 0$ becomes a rough approximation as the gravitational influence of the secondary black hole becomes effective. Nevertheless, to simplify our analysis, we adopt $u^z = 0$ for all mass ratios. Furthermore, the metric exhibits time translation ($t$) and azimuthal ($\phi$) symmetries. Consequently, the BBH metric is associated with two Killing vectors along the $t$ and $\phi$ directions. Any matter living in this background must also adhere to these symmetries to maintain the stability of the entire system. This ensures the conservation of energy and the angular momentum flux associated with the flow. Under these circumstances, the matter near the primary black hole must have angular momentum aligned with the $Z$-axis, and hence matter will approximately lie on the $z = 0$ plane as long as the mini disk is concerned. Additionally, we assume that the flow is steady ($\partial Q/\partial t = 0$), axisymmetric ($\partial Q/\partial \phi = 0$), and inviscid.

\subsection{Governing equations for CPD}
The fundamental governing equations that describe the flow motion in the CPD are given by \cite[]{Dihingia-2018-083004},
\begin{itemize}
	\item[(a)] Radial momentum equation:
	\begin{equation}
		\label{eq:radial-momentum-eq}
		\gamma_{v}^{2}v\frac{dv}{dx} + \frac{1}{e+p}\frac{dp}{dx} + \left(\frac{d\Phi^{\rm eff}}{dx}\right)_\lambda = 0,
	\end{equation}
	\item [(b)] Energy equation:
	\begin{equation}
		\label{eq:energy-eq}
		\frac{e+p}{\rho}\frac{d\rho}{dx} - \frac{de}{dx} = 0,
	\end{equation}  
\end{itemize}
where $v$ is the radial-component of the physical three-velocity in the co-rotating frame (CRF, i.e.,  a frame orbits with the same angular velocity $\Omega$ as the fluid) \cite[]{Lu-1985-176, Peitz-1996-681}, which takes negative values for accretion, and $\gamma_{v} = 1/\sqrt{1 - v^2}$ denotes the Lorentz-factor corresponding to $v$. Here, $\rho$ is the rest-mass density, $e$ ($= \rho + \Pi$, $\Pi$ is the internal energy density) is the total energy density, and $p$ is the isotropic pressure of the flow. The quantity $\Phi^{\rm eff}$ is called the effective potential of the accreting system, and it is obtained as,
\begin{equation}
	\label{eq:eff-potential}
	\Phi^{\rm eff} = 1 + \frac{1}{2}\ln\biggl[\frac{x^2V}{x^2 - \lambda^2V^2}\biggr],
\end{equation}
where $\lambda$ ($= - u_\phi/u_t$, $u_{\phi}$ and $u_t$ denote the $\phi$ and $t$ components of the covariant four-velocity $u_k$, respectively) is the specific angular momentum of the flow. Note that in the energy equation (\ref{eq:energy-eq}), we do not consider any radiative cooling processes due to the low mass accretion rate of the supermassive black holes \cite[]{Yarza-2020-50, Yuan-2014-529}.

From the time translation and azimuthal symmetries of the spacetime, we derive two conserved quantities along the stream lines of flow:
\begin{equation}
	\label{eq:conservation-E-L}
	E = -\frac{e + p}{\rho}u_t, \hspace{0.2cm} \mathcal{L} = \frac{e + p}{\rho}u_\phi,
\end{equation}
where $E$ is known as the Bernoulli constant and $\mathcal{L}$  represents the bulk angular momentum per unit mass. Therefore, $\lambda$ ($ = \mathcal{L}/E$) remains conserved along the flow direction. Using the normalization condition ($u^k u_k = -1$) of the four-velocity, we determine $u_t$ as,
\begin{equation}
	\label{eq:u_t}
	\begin{split}
		u_{t} =  - \gamma_{v} \sqrt{\frac{V}{1 - \Omega \lambda}},
	\end{split}
\end{equation}
where $\Omega$ ($= u^\phi/u^t = \lambda V^2/x^2$) refers the angular velocity of the flow.

The mass accretion rate of the flow is derived by integrating the continuity equation $\nabla_k(\rho u^k) = 0$, and it is given by,
\begin{equation}
	\label{eq:mass_accretion}
	\dot{M} = -4\pi \rho H v \gamma_{v} \sqrt{f_0},
\end{equation}
where $H$ is the local half-thickness of the disc. The expression of $H$ is obtained by considering the hydrodynamic equilibrium along the transverse direction of the disc as \cite[]{Lasota-1994-341, Riffert-1995-508, Peitz-1996-681},
\begin{equation}
	\label{eq:H}
	H = \sqrt{\frac{px^3}{\rho F}}, \hspace{0.2cm} F = \frac{1}{1 - \Omega \lambda}.
\end{equation}

To solve the dynamical equations of the flow, we require a relativistic equation of state (REoS) that relates various thermodynamic quantities of the flow (e.g., mass density, pressure, and temperature). We assume the fluid system consists of fully ionized plasma, which means fluid satisfies the charge neutrality condition $n_p = n_e \approx \rho/m_p$, where $n_p$ and $n_e$ represent the number densities of protons and electrons, respectively. Since for hot accretion flow (HAF), the plasma temperature can reach up to $\sim 10^{10-12}$K\footnote{Note that the temperature at the inner edge of the disc exceeds the typical threshold for nuclear fusion in stellar cores, which lies in the range of $10^7 - 10^8$ K. However, the mass density of matter in the accretion disc is significantly lower than that in stellar cores. Moreover, the interaction time between nuclei is very short due to the short accretion timescale. For these reasons, despite the high temperature at the disc's inner edge, the other physical conditions are not favorable for nuclear fusion to occur in the accretion disc.} \cite[]{Yuan-2014-529} near the horizon, we employ the relativistic EoS with variable adiabatic index $\Gamma$, as proposed in \cite{Chattopadhyay-2009-492}. Accordingly, the internal energy density $e$ and isotropic pressure $p$ are defined as,
\begin{equation}
	\label{eq:REoS}
	e = \frac{\rho f}{1 + m_p/m_e}, \hspace{0.2cm} p = \frac{2\rho\Theta}{1 + m_p/m_e}, 
\end{equation}
where $m_p$ and $m_e$ are masses of proton and electron, respectively, and $\Theta$ [$= k_{B}T/(m_{e}c^{2})$, $k_{B}$ is the Boltzmann constant and $T$ is the flow temperature in Kelvin] is the dimensionless temperature. Here, the quantity $f$ is given by,
\begin{equation}
	\label{eq:f}
	f = 1 + \frac{m_p}{m_e} + \Theta\left[\left(\frac{9\Theta + 3}{3\Theta + 2}\right) + \left(\frac{9\Theta + 3m_{p}/m_{e}}{3\Theta + 2m_{p}/m_{e}}\right)\right].
\end{equation}

Integrating Eq. (\ref{eq:energy-eq}) by using Eqs. (\ref{eq:REoS}) and (\ref{eq:f}), the expression of mass density $\rho$ in terms of $\Theta$ is calculated as,
\begin{equation}
	\label{eq:density}
	\rho = \mathcal{K}\exp{(\chi)}\Theta^{3/2}(3\Theta + 2)^{3/4}(3\Theta+2m_p/m_e)^{3/4}, 
\end{equation}
where $\mathcal{K}$ is the entropy constant and $\chi = (f - 1 - m_p/m_e)/(2\Theta)$. We determine the entropy accretion rate, which is also a constant of motion due to the adiabatic nature of the flow, and is given by \cite[]{Chattopadhyay-2016-3792, Kumar-2017-4221},
\begin{equation}
	\label{eq:entropy-accretion-rate}
	\begin{split}
		\mathcal{\dot{M}} = \frac{\dot{ M}}{4\pi\mathcal{K}} & = - v\gamma_{v}H\sqrt{f_0}\exp{(\chi)}\\
		& \times \Theta^{3/2}(3\Theta + 2)^{3/4}(3\Theta+2m_p/m_e)^{3/4}.
	\end{split}
\end{equation}

In the steady state, $\dot{M}$ is typically treated as a constant of motion. Therefore, we solve the equation $d\dot{M}/dx = 0$ with the help of Eqs.~(\ref{eq:energy-eq}) and (\ref{eq:REoS}) which yields the temperature gradient as,
\begin{equation}
	\label{eq:temperature-gradient}
	\frac{d\Theta}{dx} = - \frac{2\Theta}{2N + 1}\left[\frac{\gamma_v^2}{v}\frac{dv}{dx} + N_{11} + N_{12}\right],
\end{equation}
with
\begin{equation}
	N_{11} = \frac{3}{2x} - \frac{1}{2F}\frac{dF}{dx}, \hspace{0.2cm} N_{12} = \frac{1}{2f_0}\frac{df_0}{dx},
\end{equation}
where $N$ ($= \frac{1}{\Gamma - 1} = \frac{1}{2}\frac{df}{d\Theta}$) refers the polytropic index. Note that, for simplicity, the ideal equation of state (IEoS) is often used, where $N$ (or $\Gamma$) is treated as a constant. However, in our case, it is a variable quantity, as the temperature exceeds $\sim 10^{10}$ K near the horizon \cite{Chattopadhyay-2009-492, Dihingia-2019-3209}. Using the above definition of $N$, one can obtain its profile provided the temperature profile is known, which can be found from Eq. (\ref{eq:temperature-gradient}).

Subsequently, by solving Eqs. (\ref{eq:radial-momentum-eq}), (\ref{eq:energy-eq}), (\ref{eq:REoS}), and (\ref{eq:temperature-gradient}), we derive the radial-velocity gradient as,
\begin{equation}
	\label{eq:velocity-gradient}
	\frac{dv}{dx} = \frac{\mathcal{N}}{\mathcal{D}}.
\end{equation}
The expressions of numerator ($\mathcal{N}$) and denominator ($\mathcal{D}$) are given by,
\begin{align}
	\mathcal{N} & = \frac{2C_{s}^2}{\Gamma + 1} (N_{11} + N_{12}) - \frac{d\Phi^{\rm eff}}{dx},\\
	\mathcal{D} & = \gamma_{v}^{2}\left[v - \frac{2C_{s}^{2}}{(\Gamma + 1)v}\right],
\end{align}
where $C_s$ ($= \frac{\Gamma p}{e + p} = \frac{2\Gamma \Theta}{f + 2\Theta}$) is the adiabatic sound speed.

The above flow equations are useful for determining the accretion solutions in the CPD. However, we mentioned earlier that in the absence of both the external field and the secondary black hole, the spacetime metric of a BBH system [Eq. (\ref{eq:BBH-metric})] reduces to that of a single Schwarzschild black hole of mass $m_1$ (see Eqs. (\ref{eq:deformed-Sch-metric}) and (\ref{eq:deformed-Sch-metric-1})). Consequently, all the flow equations, namely Eqs. (\ref{eq:eff-potential}), (\ref{eq:u_t}), (\ref{eq:mass_accretion}), (\ref{eq:entropy-accretion-rate}), and (\ref{eq:temperature-gradient}), must reduce to the standard flow equations corresponding to an isolated Schwarzschild black hole. Therefore, one can expect that these equations provide accretion solutions similar to those reported in the literature for an isolated Schwarzschild black hole \cite[and references therein]{Chattopadhyay-2009-492, Chattopadhyay-2016-3792}.

\subsection{Conditions for critical point}
\label{sec:critical-point-conditions} 
In case of transonic accretion flow, it must cross the event horizon ($x_{\rm H}$) with radial velocity close to the speed of light ($v \approx 1$), thereby satisfying the inner boundary condition at the event horizon \cite[]{Liang-1980-271, Abramowicz-1981-314}. Usually, accreting matter begins its journey with negligible radial velocity ($v << 1$) at the outer edge of the accretion disc. As the flow moves inward, it must pass through at least one critical point ($x_c$) where it changes the sonic state from subsonic to supersonic. At $x_c$, the radial velocity gradient takes the indeterminate form $(dv/dx)_{x_c} = 0/0$. Therefore, for the critical points, we find the following conditions:
\begin{equation}
	\label{eq:critical-point-conditions}
	\mathcal{N} = \mathcal{D} = 0.
\end{equation} 

To calculate the real and finite values of $(dv/dx)_{x_c}$, we apply l$'$H\^{o}pital's rule to Eq. (\ref{eq:velocity-gradient}). We obtain two distinct values of $(dv/dx)_{x_c}$, and depending on them, critical points are classified into three categories: saddle, nodal, and spiral types. If both values of $(dv/dx)_{x_c}$ are real with opposite sign, the critical point is called as a saddle-type critical point. For a nodal-type critical point, $(dv/dx)_{x_c}$ values are real and have same sign. When $(dv/dx)_{x_c}$ is imaginary, the critical point is classified as a spiral-type critical point. It is noted that a positive value of $(dv/dx)_{x_c}$ yields the accretion solution, while a negative value of $(dv/dx)_{x_c}$ gives the wind solution. In this work, we focus only on the accretion solutions, leaving the wind solutions for future study. Additionally, among the various types of critical points, saddle-type critical points are physically acceptable. Therefore, we concentrate on the accretion solutions that only pass through the saddle-type critical points. We further point out that the flow can have single or multiple critical points depending on the model parameters. A detailed analysis of the critical points is discussed in Section~\ref{sec:critical-points-analysis}. 

The above equations are essential for determining the accretion solutions and their associated thermodynamic flow variables. In this study, we also aim to explore the spectral characteristics of the accretion solutions in terms of emissivity, luminosity distribution, and optical depth, etc. The required model equations for this analysis are presented in the following section. 

\subsection{Radiative properties of accretion flow}
\label{eq:radiative-properties}
We consider the relativistic thermal bremsstrahlung (free-free) emission from the CPD. For HAF, electron-electron emission dominates over electron-ion emission \cite{Patra-2023-060}, and hence we use an approximate expression for the free-free emissivity, given by \cite[]{Novikov-1973-343},

\begin{equation}
	\label{eq:emissivity}
	\begin{split}
		\mathcal{E}_{\nu_e}^{\rm ff} & = \frac{32\pi e^{6}}{3m_{e}m_p^2c^{3}}\sqrt{\frac{2\pi}{3m_{e}k_{B}}}Z^2\rho^2T_{e}^{-1/2}(1 + 4.4\times 10^{-10}T_{e})
		\\& \times \exp\biggl(-\frac{h\nu_e}{k_{B}T_{e}}\biggr)\bar{g}_{\text{ff}}~{\text{erg}~\text{s}^{-1}~\text{cm}^{-3}~\text{Hz}^{-1}},
	\end{split}
\end{equation}
where $Z$ represents the atomic number of the ion, which is taken as $1$ for the hydrogen plasma, $h$ is the Planck constant, and $\nu_e$ is the emission frequency. The term $\bar{g}_{\rm ff}$ denotes the thermally-averaged Gaunt factor, which does the quantum mechanical correction to the classical electrodynamics. Its value varies between $1$ to $5$, depending on the electron energy. However, following \cite{Yarza-2020-50}, we take $\bar{g}_{\rm ff} = 1.2$ throughout this paper. Note that the second term in Eq.~(\ref{eq:emissivity}) includes both the relativistic corrections and electron-electron bremsstrahlung emission. We also assume a two-temperature plasma in which the electron temperature is lower than that of the ions. Here, we briefly discuss the physical reasons behind this assumption. In an ionized plasma, electrons and ions attempt to maintain thermal equilibrium through random Coulomb collisions. However, this equilibrium breaks down, particularly near the horizon. In this region, the flow velocity becomes extremely high, and the accretion timescale becomes much shorter than the electron-ion collision timescale. As a result, the assumption of a single-temperature plasma becomes unrealistic, and the accretion flow tends to enter a two-temperature regime, especially in the inner regions of the disc. There are many studies in the literature corresponding to the two-temperature accretion flows \cite[]{Sarkar-2018-1950037, Dihingia-2018-1, Dihingia-2020-3043, Sarkar-2020-A209, Sarkar-2022-34}, where the electron and ion temperature profiles are obtained self-consistently by solving the radial momentum and energy equations for each species. However, in our work, we adopt a simplified scaling relation, $T_e = T/10$ in accordance with the works \cite{Monika-2009-497, Yarza-2020-50}. In those studies, the authors employed this scaling relation to successfully model the observed data of Sgr A*. However, readers may adopt some alternative scaling relations as discussed in works \cite{Chattopadhyay-2002-454, Dihingia-2020-023012}.

The emission coefficient of accretion flow, as described in Eq. (\ref{eq:emissivity}), is measured locally. However, for an observer at spatial infinity, this radiation must be red-shifted due to both the gravitational potential of the black hole and the Doppler effect resulting from the rotation of the disc. Here, we do not consider the light bending effect to the photons for simplicity. Also, the disc inclination angle with respect to the line of sight of the observer is taken as $45^{\circ}$ \cite[]{Dihingia-2020-023012, Sen-2022-048, Patra-2023-060}. Under these assumptions, the observed frequency $\nu_o$ is obtained in terms of the emitted frequency $\nu_e$ as \cite[]{Luminet-1979-228},
\begin{equation}
	\label{eq:red-shift}
	\nu_o = \frac{\nu_e}{1+ z} =  \frac{\nu_e}{u^{t}\left(1 + \frac{r\Omega}{\sqrt{2}c}\sin{\phi}\right)},
\end{equation}
where ($1 + z$) is the red-shift factor. The expression of $u^t$ is calculated from Eq. (\ref{eq:u_t}), and is given by,
\begin{equation}
	\label{eq:u_upper_t}
	u^{t} = \frac{\gamma_{v}}{\sqrt{(1 - \Omega \lambda)V}}.
\end{equation}

Using Eqs. (\ref{eq:emissivity}) and (\ref{eq:red-shift}), we get the monochromatic luminosity of the CPD at a observed frequency $\nu_o$ as,
\begin{equation}
	\label{eq:monochromatic-luminosity}
	\begin{split}
		L_{\nu_o} & = 2\int_{x_0}^{x_{\rm edge}}\int_{0}^{2\pi}\mathcal{E}_{\nu_{e}}^{\rm ff}Hxdx d\phi
		\\& = \frac{64\pi e^{6}}{3m_{e}m_p^2c^{3}}\sqrt{\frac{2\pi}{3m_{e}k_{B}}}\bar{g}_{\rm ff} \int_{x_0}^{x_{\rm edge}}\int_{0}^{2\pi}\biggl[Hx\rho^{2}T_{e}^{-1/2}
		\\& \times (1 + 4.4\times 10^{-10} T_{e})\exp\left(-\frac{(1+z)h\nu_o}{k_{B}T_e}\right)\biggr]dx d\phi
		\\& \times {\text{erg}~\text{s}^{-1}~\text{Hz}^{-1}} , 
	\end{split}
\end{equation}
where $x_0$ ($= x_{\rm H} =$ event horizon radius) and $x_{\rm edge}$ are the inner and outer edges of the CPD, respectively.

Bremsstrahlung radiation can only be observed when the emitting medium is optically thin; otherwise, photons will be destroyed through true absorption after undergoing multiple coherent scattering. The primary source of opacity is the Thomson scattering by the free electrons with opacity coefficient $\kappa_s = 0.4~{\text{cm}^2~\text{gm}^{-1}}$ \cite[]{Maoz-2016-book}. In a fully ionized medium, free-free absorption also significantly contributes to the photon opacity \cite[]{Maoz-2016-book}. Using Eq. (\ref{eq:emissivity}), we calculate the Rosseland mean opacity coefficient for the free-free absorption as \cite[]{Shapiro-2008-book},
\begin{equation}
	\label{eq:absorption-opacity-coefficient}
	\begin{split}
		\kappa_{\rm R}^{\text{ff}} & = 0.64\times 10^{23}\rho T_{e}^{-7/2}\\
		& \times (1 + 4.4\times 10^{-10}T_{e})\bar{g}_{\text{ff}}~{\text{cm}^{2}~\text{gm}^{-1}}.
	\end{split}
\end{equation} 
Here, we consider the typical length scale of the medium is the half-thickness ($H$) of the disc. Therefore, the effective optical depth of the disc is obtained by considering the aforementioned opacity sources as \cite[]{Rybicki-1991-book},
\begin{equation}
	\label{eq:effective-optical-depth}
	\tau_{\text{eff}} \approx \sqrt{\tau_{a}(\tau_{a} + \tau_{s})},
\end{equation}
where $\tau_s$ ($= \kappa_{s}\rho H$) is the scattering optical depth and $\tau_a$ ($= \kappa_{\rm R}^{\text{ff}}\rho H$) is the absorption optical depth.

With these useful equations, we present the results regarding the accretion properties of the CPD in the subsequent sections.

\section{Results}
\label{sec:results}

\subsection{Analysis of critical points}
\label{sec:critical-points-analysis}

\begin{figure}
	\centering
	\includegraphics[width=\columnwidth]{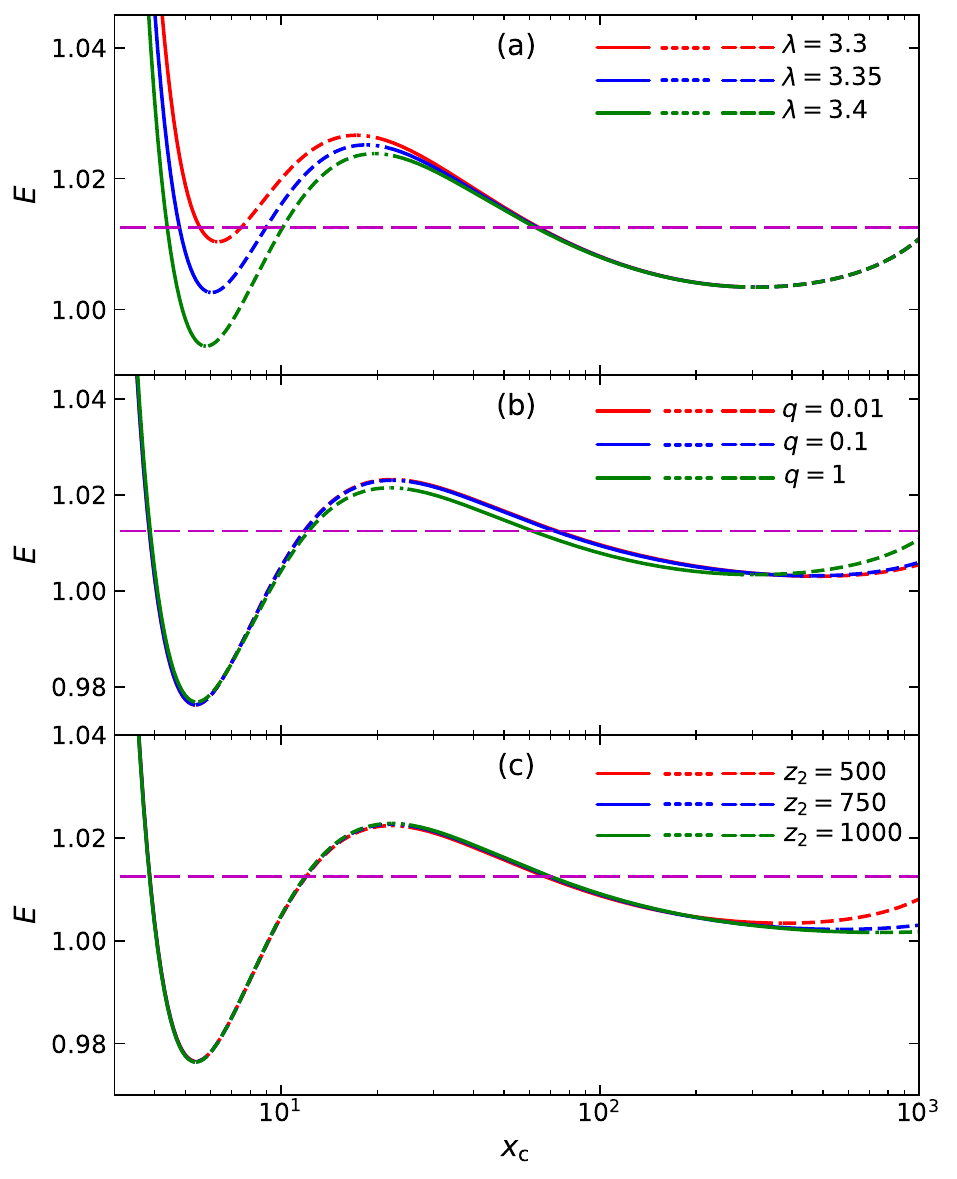}
	\caption{Plot of specific energy ($E$) as a function of critical point location ($r_c$) for different angular momentum ($\lambda$) in panel (a), binary mass ratio ($q$) in panel (b), and binary separation ($z_2$) in panel (c). Here, solid, dotted, and dashed curves represent saddle, nodal, and spiral types critical points, respectively. The horizontal dashed lines (magenta) denote the energy values where the flow possesses multiple critical points. See the text for details.}
	\label{fig:E-r}
\end{figure}

We already pointed out that the transonic accretion flow must pass through the critical points; therefore, we start our analysis by investigating how the behaviors of critical points depend on different global constants, such as specific angular momentum ($\lambda$), binary mass ratio ($q$), and binary separation ($z_2$). To do this, we first compute the flow temperature ($\Theta_c$) and radial velocity ($v_c$) at a critical point ($x_c$) for a given set of input parameters ($\lambda, q, z_2$) by solving the critical point conditions (\ref{eq:critical-point-conditions}). Subsequently, using these results in Eq. (\ref{eq:conservation-E-L}), we calculate the specific energy ($E$) at $x_c$ for the same set of input parameters mentioned above. The obtained results are shown in Fig. \ref{fig:E-r}, where the variation of $E$ as a function of $x_c$ is illustrated. In panel (a), we fix $(q, z_2) = (1, 500)$ and vary the angular momentum as $\lambda = 3.3, 3.35$, and $3.4$. In panel (b), we set $(\lambda, z_2) = (3.5, 500)$ and vary the mass ratio as $q = 0.01, 0.1$, and $1$. Finally, in panel (c), we choose $(\lambda, q) = (3.5, 0.5)$ and vary the binary separation as $z_2 = 500, 750$, and $1000$. The color codes corresponding to the different input parameters are indicated in each panel. In each curve, the saddle, nodal, and spiral-type critical points are represented by solid, dotted, and dashed lines, respectively. We observe that the flow energy remains constant in all cases when the critical points are located at large distances from the black hole. However, the flow energy changes significantly when the critical points are closer to the black hole. For each curve, the critical points are formed in a systematic order as saddle-nodal-spiral-nodal-saddle-spiral when the critical points move toward the outer edge of the disk. Moreover, we observe that the flow energy changes moderately with an increase in $\lambda$. However, it shows small variation with increases in $q$ and $z_2$. Interestingly, a broad range of energy values exists where the flow possesses two saddle-type critical points (as indicated by the horizontal dashed lines (magenta) in all the panels). It is important to note that such regions with multiple critical points are of particular interest in the study of shock-induced accretion solutions. For the shock scenarios \cite[]{Dihingia-2018-083004, Dihingia-2019-2412, Patra-2022-101120, Sen-2022-048, Mitra-2024-28, Patra-2024-371}, a global solution can pass through both the inner (i.e., formed near the black hole) and outer (i.e., formed far from the black hole) critical points, provided that the relativistic shock conditions are satisfied. A detailed analysis of the shock solutions is presented in Section \ref{sec:shock-solutions}. 

\subsection{Effect of binary parameters on the transonic accretion solutions}
\label{sec:transonic-solutions} 

\begin{figure}
	\centering
	\includegraphics[width=\columnwidth]{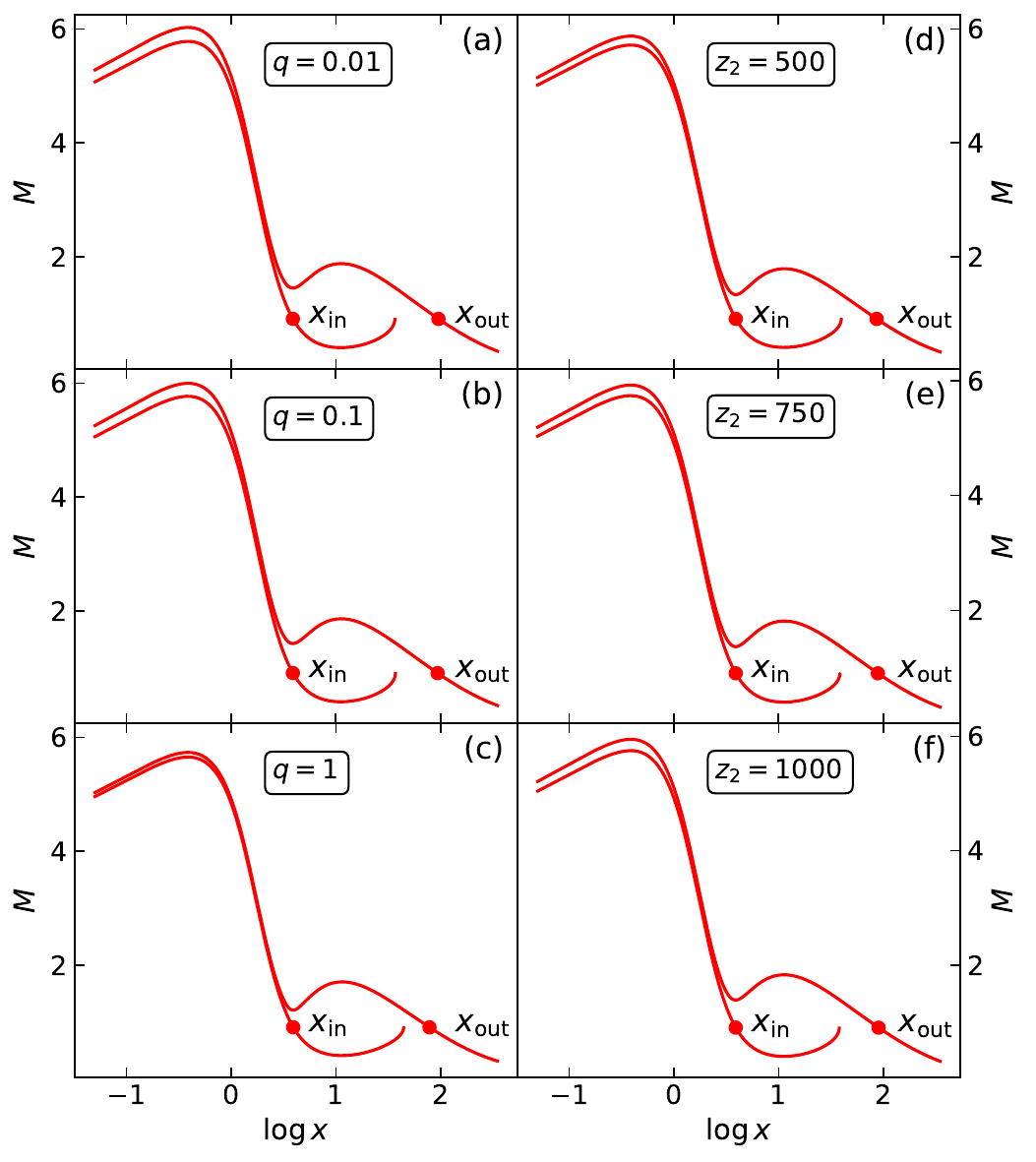}
	\caption{Plot of Mach number ($M = v/C_s$) as a function of radial distance ($x$) for the binary mass ratio $q = 0.01$, $0.1$, and $1$ with binary separation $z_2 = 500$ (left panels), and for $z_2 = 500$, $750$, and $1000$ with $q = 0.5$ (right panels). Here, the critical points are marked by the filled circles, and we choose $\lambda = 3.5$ and $E = 1.01$. See the text for details.}
	\label{fig:solutions}
\end{figure}  

\begin{table}
	\centering
	\caption{Binary mass ratio ($q$), binary separation ($z_2$), inner critical point ($x_{\rm{in}}$), outer critical point ($x_{\rm{out}}$), and terminate radius ($x_t$) for accretion solutions presented in Fig.~\ref{fig:solutions}.}
	\label{tab:table-1}
	\begin{ruledtabular}
		\begin{tabular}{lcccc}
			$q$ & $z_2$ & $x_{\rm{in}}$ & $x_{\rm{out}}$ & $x_t$\\
			\hline
			0.01 & 500 & 3.8899 & 95.5954 & 36.6399\\
			0.1 & 500 & 3.8924 & 93.8551 & 37.1224\\
			1 & 500 & 3.9180 & 78.4602 & 44.5380\\
			0.5 & 500 & 3.9038 & 86.5751 & 39.7038\\
			0.5 & 750 & 3.8990 & 88.8785 & 38.5990\\
			0.5 & 1000 & 3.8967 & 90.2864 & 38.0367\\
		\end{tabular}
	\end{ruledtabular}
\end{table}  

In this section, we aim to investigate the effect of the secondary black hole (SBH) on the accretion solutions in the CPD. To determine the transonic accretion solutions, we calculate the critical point $x_c$ and its corresponding flow variables $v_c$ and $\Theta_c$ for a given set of input parameters ($q, z_2, \lambda, E$) by using Eqs. (\ref{eq:critical-point-conditions}) and (\ref{eq:conservation-E-L}). Next, using the calculated parameters ($\Theta_c, v_c$) at $x_c$ as the initial boundary conditions, we first integrate Eqs. (\ref{eq:temperature-gradient}) and (\ref{eq:velocity-gradient}) from $x_c$ to the disc inner edge $x_0$, and then from $x_c$ to the disc outer edge $x_{\text{edge}}$. Finally, by combining both segments of the solutions, we obtain the resulting global accretion solution corresponding to the same set of input parameters ($q, z_2, \lambda, E$). In this work, the inner and outer edges of the accretion disc are taken as $x_0 = 0.05$ and $x_{\text{edge}} = 350$, respectively. It is essential to mention that there is no qualitative difference in the characteristics of the accretion solutions if the inner edge position is chosen more closer to the horizon ($x_{\rm H} = 0$) instead of $x_0 = 0.05$. In Fig. \ref{fig:solutions}, we depict the typical accretion solutions (i.e., variation of the Mach number ($M = v/C_s$) as a function of the radial coordinate ($x$)) for different values of  $q$ and $z_2$. Here, we choose the input flow parameters as $\lambda = 3.5$ and $E = 1.01$. The left panels represent the accretion solutions corresponding to the mass ratio $q =0.01, 0.1$, and $1$ with binary separation $z_2 = 500$. Similarly, the right panels denote the accretion solutions for $z_2 = 500, 750$, and $1000$ with $q = 0.5$. In all cases, the solutions that pass through the outer critical points ($x_{\text{out}}$) are globally extended from $x_{\text{edge}}$ to $x_{0}$. In contrast, the solutions that pass through the inner critical points ($x_{\text{in}}$) are truncated at certain radii ($x_t$) in between $x_{\text{in}}$ and $x_{\text{out}}$. Accretion solutions of this kind are usually referred as A-type solutions \cite[]{Dihingia-2018-083004, Dihingia-2020-023012, Sen-2022-048, Patra-2022-101120, Patra-2023-060, Patra-2024-371}. Interestingly, for A-type solution topology, the solution that passes through $x_{\text{out}}$ may connect to the solution that passes through $x_{\text{in}}$ via a standing shock transition, as the inner solution branch has a higher entropy content than the outer one \cite[]{Becker-2001-429}. The shock-induced accretion solutions will be examined latter in Section \ref{sec:shock-solutions}. It is evident from the figure that the nature of the accretion solutions does not change with an increase in either of the binary parameters $q$ and $z_2$. The positions of the critical points associated with these accretion solutions are summarized in Table \ref{tab:table-1}. This table shows that $x_{\text{in}}$ increases, while $x_{\text{out}}$ decreases as $q$ increases at a given $z_2$. On the other hand, when $z_2$
increases at a fixed $q$, $x_{\text{in}}$ and $ x_{\text{out}}$ move oppositely compared to the previous case.

\subsection{Physical properties of the CPD with shock}
\label{sec:shock-solutions} 

\begin{figure*}
	\centering
	\includegraphics[width=0.7\linewidth]{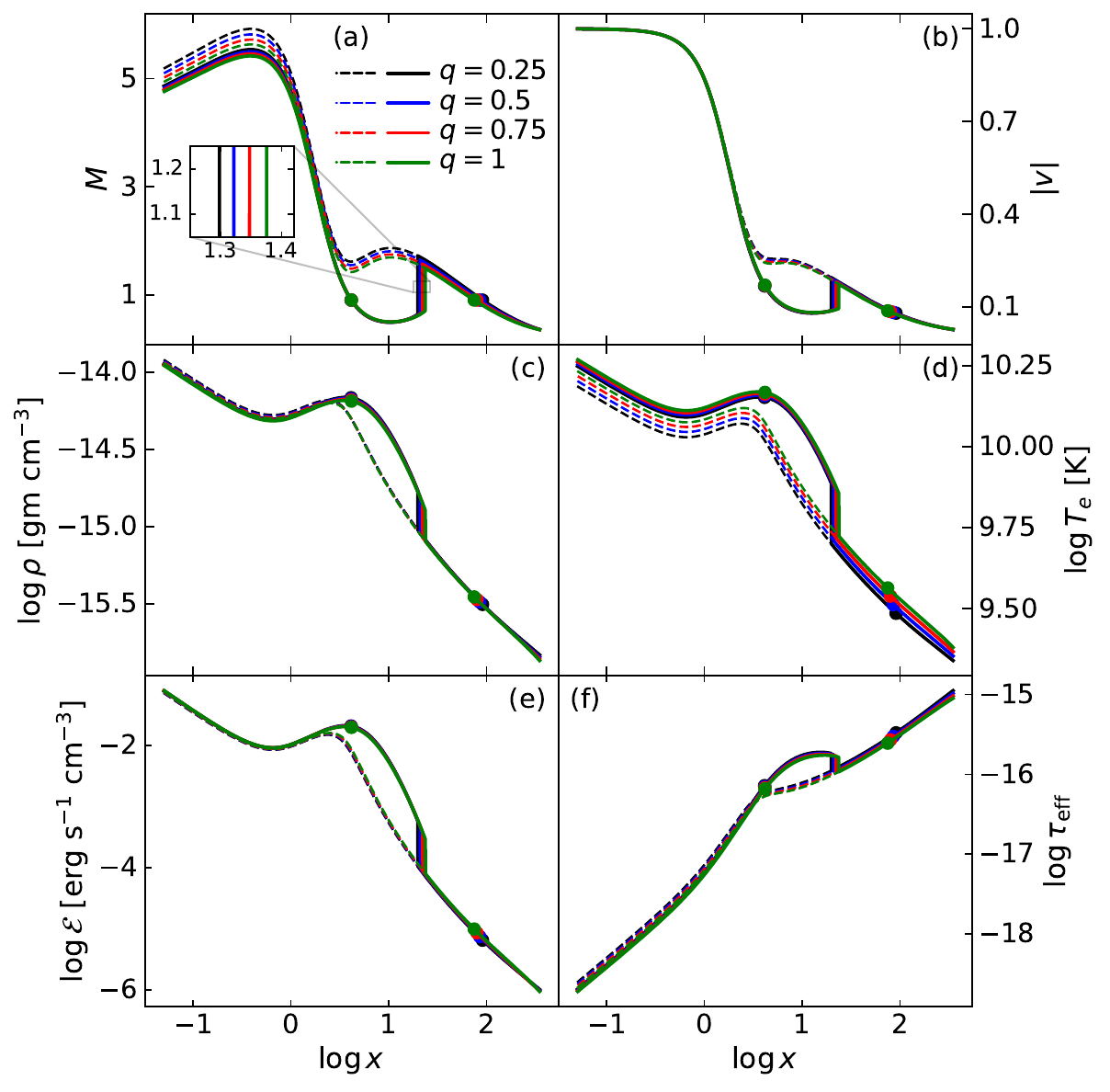}
	\caption{Illustration of the shock-induced accretion solutions (Mach number ($M$) versus radial distance ($x$) curves (solid) in panel (a)) for binary mass ratios $q = 0.25$, $0.5$, $0.75$, and $1$ with binary separation $z_2 = 350$. The respective  profiles of radial-velocity ($v$) in panel (b), mass density ($\rho_0$) in panel (c), electron temperature ($T_e$) in panel (d), frequency-integrated emissivity ($\mathcal{E}$) in panel (e), and effective optical depth ($\tau_{\text{eff}}$) in panel (f). The dashed curves denote the scenario where shock transitions have not occurred. The flow parameters for this figure are chosen as $\lambda = 3.45$ and $E = 1.01$. See the text for details.}
	\label{fig:sh-solutions-diff-q}
\end{figure*}

In this section, we focus on studying the effect of binary parameters ($q, z_2$) on the shock-induced accretion solutions and examine their corresponding physical properties. In Fig. \ref{fig:sh-solutions-diff-q}a, we present the dynamical structure of the shock solutions for different values of mass ratio ($q$) with binary separation $z_2 = 350$, where the Mach number ($M$) is plotted as a function of radial distance ($x$). Here, we choose $\lambda = 3.45$ and $E = 1.01$. The solid curves with black, blue, red, and green colors denote the shock solutions for $q = 0.25, 0.5, 0.75$, and $1$, respectively. However, the dashed curves with the same color code represent the respective shock-free solutions. The locations of the critical points ($x_{\text{in}}, x_{\text{out}}$) and shock radius ($x_{\text{sh}}$) for these shock solutions are tabulated in Table \ref{tab:table-2}. In a general scenario, the accretion solution, after passing through the outer critical point $x_{\text{out}}$, continues to move toward the central black hole until it crosses the event horizon (see the dashed curves). Meanwhile, the flow slows down when the centrifugal repulsion due to flow rotation becomes dominant against the gravitational pull of the central black hole. Consequently, this supersonic flow jumps into the subsonic branch, which contains $x_{\text{in}}$. Thereafter, the flow becomes supersonic once more after crossing $x_{\text{in}}$, and continues to accrete toward the horizon. Therefore, in the shock scenario, the accretion solution can pass through both $x_{\text{in}}$ and $x_{\text{out}}$ via a discontinuous jump in between them. To calculate the shock location, we use the following relativistic shock conditions which are given by \cite{Taub-1948-328},
\begin{itemize}
	\item[(a)] Mass flux conservation: $[\rho u^x]$,
	\item [(b)] Energy flux conservation: $[(e + p)u^{x}u^{t}]$,
	\item [(c)] Radial-momentum flux conservation:\\ $[(e + p)u^{x}u^{x} + pg^{xx}]$,
\end{itemize}
where the quantities with the square bracket denote the difference of their values across the shock front. Using these shock conditions, we calculate the shock locations at $x_{\text{sh}} = 19.8524, 20.9639, 22.2453$, and $23.7268$ for $q = 0.25, 0.5, 0.75$, and $1$, respectively, as indicated by the solid vertical lines. It is observed that the shock location moves away from the horizon as the mass ratio increases. In the respective panels (b)–(f) of Fig. \ref{fig:sh-solutions-diff-q}, we present the profiles of radial-velocity ($v$), mass density ($\rho$), electron temperature ($T_e$), frequency-integrated emissivity ($\mathcal{E}$), and effective optical depth ($\tau_{\text{eff}}$) for both shock-free and shock-induced accretion solutions in Fig. \ref{fig:sh-solutions-diff-q}a. In this work, we consider a SMBHB with the primary black hole mass $m_1 = 10^6M_{\odot}$, where $M_{\odot}$ is the Solar mass. The mass accretion rate of the flow is taken as $\dot{M} = 10^{-5}\dot{M}_{\text{Edd}}$, where $\dot{M}_{\text{Edd}} = 1.39 \times 10^{18}m_1/M_{\odot}~\text{gm}~\text{s}^{-1}$ is the  Eddington mass accretion rate. We observed that all the analyzed quantities significantly change at the shock locations. Moreover, the differences in these quantities across the shock fronts diminish as the mass ratio increases. This result is expected because the shock fronts settle at larger radii as the mass ratio increases. Since the radial velocity drops down at the shock radius, according to the conservation of the mass flux across the shock front (see the shock condition (a)), the mass density of the post-shock corona (hereafter PSC) increases. Note that the mass density lies in the range of approximately $10^{-16} - 10^{-14}$ gm cm$^{-3}$. Since we have adopted a very small value of $\dot{M}$, as mentioned above, Eq. (\ref{eq:mass_accretion}) yields correspondingly low values of $\rho$. This choice of $\dot{M}$ is reasonable, as we are dealing with a radiatively inefficient accretion flow (RIAF), where $\dot{M} << \dot{M}_{\rm Edd}$. For example, in the case of Sgr A*, $\dot{M} \simeq 4.5 \times 10^{-7} \dot{M}_{\rm Edd}$, and the corresponding $\rho$ is estimated to lie in the range $10^{-18} - 10^{-14}$ gm cm$^{-3}$ \cite{Yuan-2003-301}. Another example is M87*, for which $\dot{M} \lesssim 9.2 \times 10^{-4} \dot{M}_{\rm Edd}$ \cite{Kuo-2014-L33}. Therefore, also for M87*, one might expect $\rho$ to lie within a similar range. It is worth noting that if higher values of $\dot{M}$ are considered, as suggested by some studies in the literature for Sgr A* and M87* \cite{Monika-2009-497, Yarza-2020-50}, then, according to Eq. (\ref{eq:mass_accretion}), the corresponding values of $\rho$ will increase accordingly. Since these supermassive black hole systems exhibit very low mass densities, they are optically thin, and their luminosities are low -- an important characteristic of RIAFs or hot accretion flows (HAFs) (see the review paper \cite{Yuan-2014-529}). After the shock transition, the kinetic energy of the post-shock flow is converted into the thermal energy, leading to an increase in the electron temperature of the PSC. It is important to note that the soft photons emitted by the pre-shock disk may undergo inverse Compton scattering by the swarm of hot electrons in the PSC, resulting in the production of high-energy radiations \cite[]{Roedig-2014-115}. We also observe that the emissivity of the radiation in the PSC is greater than that in the pre-shock disk due to the higher mass density and electron temperature of the flow in the PSC compared to the pre-shock disk. Furthermore, we find that the effective optical depth of the accretion disk is substantially low, even in the case of shock solutions. As a result, the disk remains optically thin, allowing the photons to escape the medium before being absorbed by the accretion disk.

\begin{table}
	\centering
	\caption{Binary mass ratio ($q$), binary separation ($z_2$), inner critical point ($x_{\rm{in}}$), outer critical point ($x_{\rm{out}}$), and shock location ($x_{\text{sh}}$) for shock-induced accretion solutions presented in Fig.~\ref{fig:sh-solutions-diff-q}.}
	\label{tab:table-2}
	\begin{ruledtabular}
		\begin{tabular}{lcccc}
			$q$ & $z_2$ & $x_{\rm{in}}$ & $x_{\rm{out}}$ & $x_{\text{sh}}$\\
			\hline
			0.25 & 350 & 4.1324 & 91.3261 & 19.8524\\
			0.5 & 350 & 4.1439 & 85.5283 & 20.9639\\
			0.75 & 350 & 4.1553 & 80.0888 & 22.2453\\
			1 & 350 & 4.1668 & 75.0221 & 23.7268\\
			
			0.5 & 300 & 4.1477 & 85.2372 & 21.2977\\
			0.5 & 400 & 4.1410 & 86.0856 & 20.7110\\
			0.5 & 500 & 4.1370 & 87.2893 & 20.3470\\
			0.5 & 600 & 4.1343 & 88.3404 & 20.1043\\
		\end{tabular}
	\end{ruledtabular}
\end{table}

\begin{figure*}
	\centering
	\includegraphics[width=0.7\linewidth]{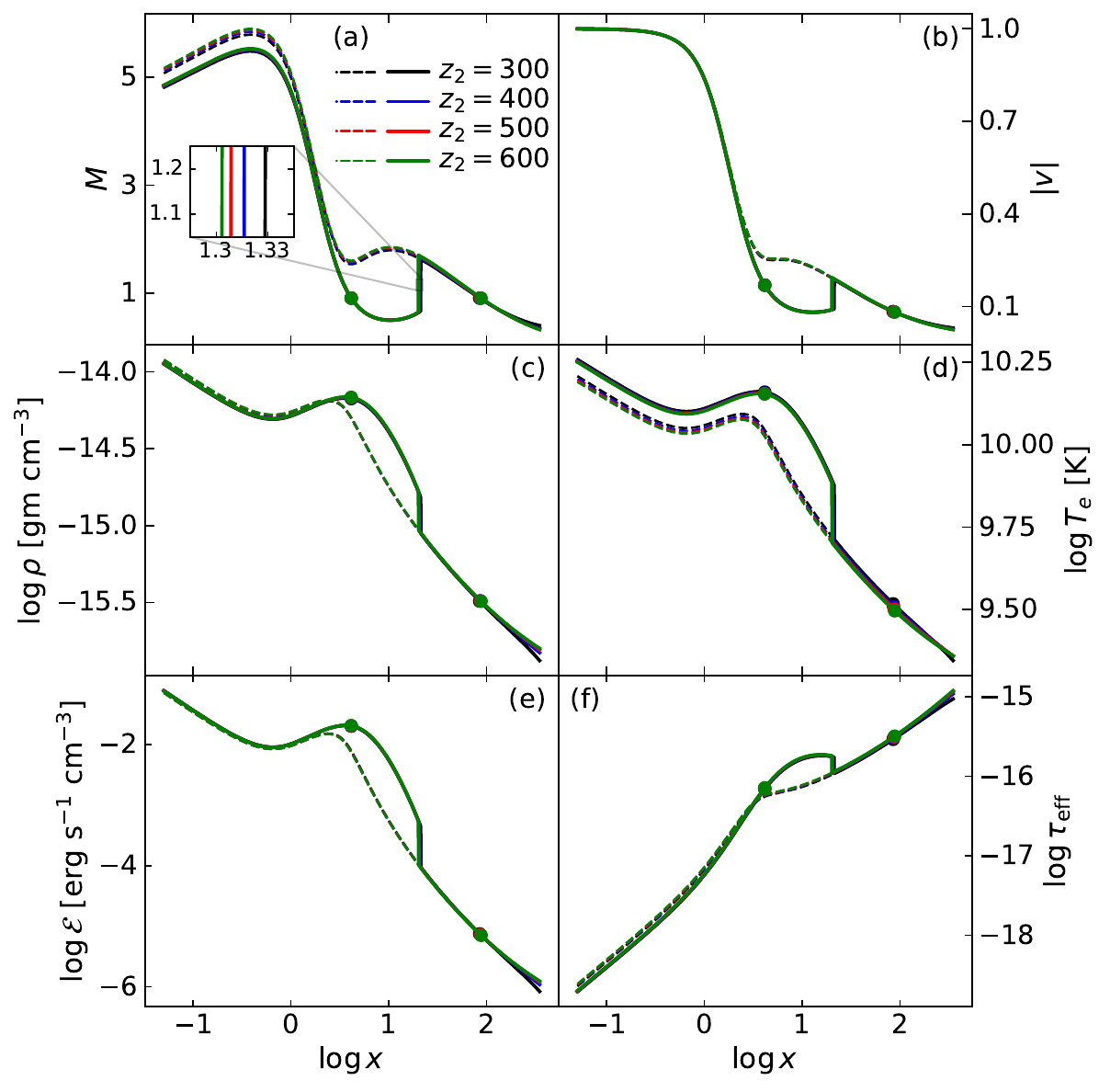}
	\caption{Typical shock solutions (Mach number ($M$) versus radial distance ($x$) curves (solid) in panel (a)) for binary separations $z_2 = 300$, $400$, $500$, and $600$ with mass ratio $q = 0.5$. The profiles of radial-velocity ($v$), mass density ($\rho$), electron temperature ($T_e$), frequency-integrated emissivity ($\mathcal{E}$), and effective optical depth ($\tau_{\text{eff}}$) are depicted in panels (b)-(f), respectively. The dashed curves represent the situations in which shock transitions have not occurred. In this figure, we choose $\lambda = 3.45$ and $E = 1.01$. See the text for details.}
	\label{fig:sh-solutions-diff-z2}
\end{figure*}

Similarly, for a fixed mass ratio $q = 0.5$ and varying binary separation $z_2$, we present the typical shock solutions in Fig. \ref{fig:sh-solutions-diff-z2}a. The corresponding flow variables such as the radial-velocity ($v$), mass density ($\rho$), electron temperature ($T_e$), frequency-integrated emissivity ($\mathcal{E}$), and effective optical depth ($\tau_{\text{eff}}$) are shown in Figs. \ref{fig:sh-solutions-diff-z2}b-f, respectively. The obtained results are denoted by the solid lines with black, blue, red, and green colors for $z_2 = 300$, $400$, $500$, and $600$, respectively. The shock-free solutions corresponding to the same set of input parameters are shown by the dashed curves with the same color codes as those used for the shock-induced solutions. For $z_2 = 300$, $400$, $500$, and $600$, we obtain the shock locations at $r_{\text{sh}} = 21.2977$, $20.711$, $20.3470$, and $20.1043$, respectively. We observe that for a given $q$, the shock radius decreases with $z_2$. As a result, the change in flow variables across the shock front increases. Here, the flow remains optically thin throughout the disc; hence, the bremsstrahlung radiation can be emitted from the accretion disc without lost its information.

\begin{figure*}
	\centering
	\includegraphics[width=0.975\linewidth]{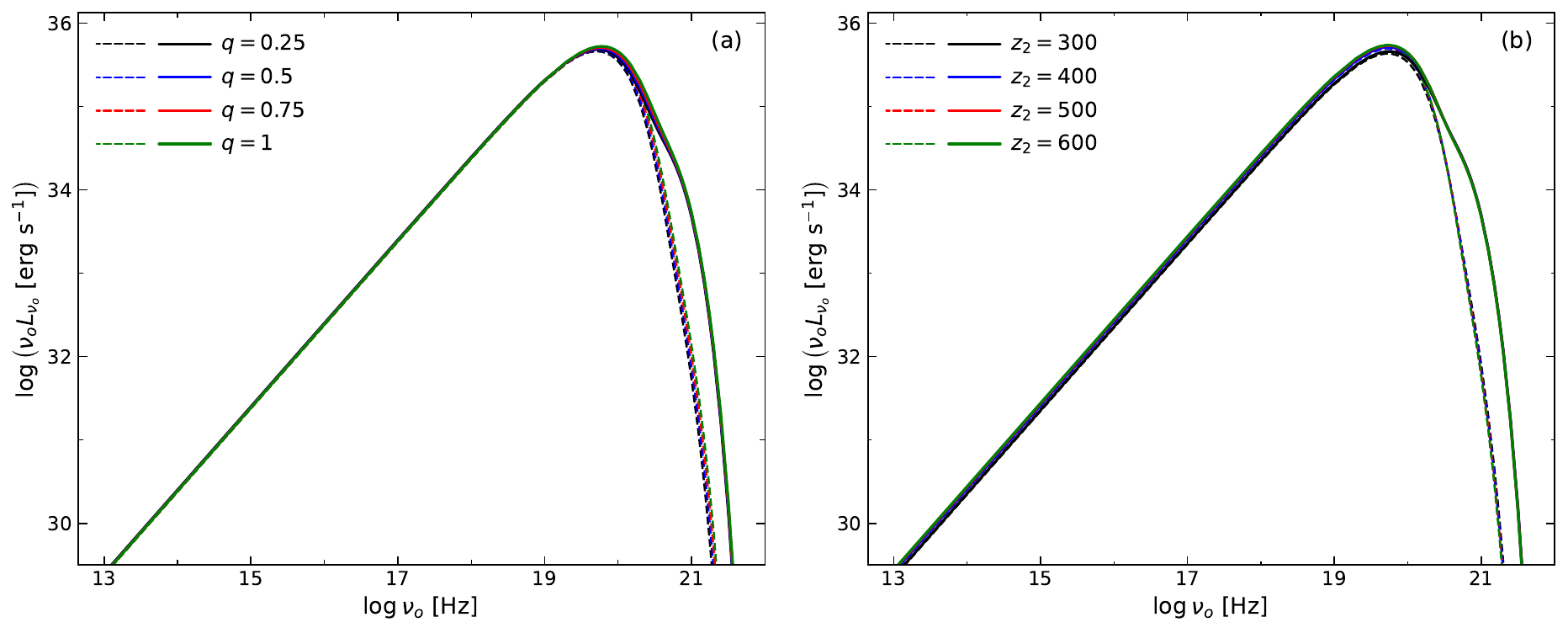}
	\caption{Spectral energy distribution (i.e., $\nu_{o}L_{\nu_{o}}$ versus $\nu_{o}$ curves) of the emitted radiation from the CPD. In panels (a) and (b), the SEDs are presented for the accretion solutions in Figs. \ref{fig:sh-solutions-diff-q} and \ref{fig:sh-solutions-diff-z2}, respectively. Here, we choose the set of input parameters as $r_{0} = 0.05$, $r_{\rm{edge}} = 350$, $\lambda = 3.45$, $E = 1.01$, $m_1 = 10^6M_{\odot}$, and $\dot{M} = 10^{-5}\dot{M}_{\text{Edd}}$. See the text for details.}
	\label{fig:SED}
\end{figure*}

Now, we investigate the spectral energy distribution (SED) of the CPD by considering the emission of thermal bremsstrahlung radiation from the accretion disc. We calculate the SED for the accretion solutions presented in Figs. \ref{fig:sh-solutions-diff-q} and \ref{fig:sh-solutions-diff-z2} using Eq. (\ref{eq:emissivity}). The results are shown in Figs. \ref{fig:SED}a-b, respectively, where the variation of the quantity $\nu_{o}L_{\nu_{o}}$ with the observed frequency $\nu_0$ is depicted. In both panels, we observe that the emitted radiation reaches its maximum power at $\nu_{o} \approx 10^{20}\text{Hz}$ \cite[]{Yarza-2020-50}. Since the electron temperature at the inner edge of the disc is $T_{e0} \approx 10^{10}\text{K}$, the spectra exhibit a cutoff at $\nu_{o} \approx 10^{21}\text{Hz}$ ($= k_BT_{e0}/h$) \cite[]{Novikov-1973-343}. It is also found that the SEDs for the shock solutions are higher than those for the shock-free solutions \cite[]{Patra-2024-371}. This occurs because the temperature distribution in the disc for a shock solution is higher, particularly in the PSC, compared to that in a shock-free solution. Furthermore, across the entire frequency domain, the CPD spectra show negligible dependence on the binary parameters $q$ and $z_2$ for both the shock-induced and shock-free solutions. This is because the electron temperature of the CPD is minimally affected by $q$ and $z_2$ (see Figs. \ref{fig:sh-solutions-diff-q}d and \ref{fig:sh-solutions-diff-z2}d).

\subsection{Shock properties of the CPD}
\label{sec:shock-properties}

\begin{figure*}
	\centering
	\includegraphics[width=0.975\linewidth]{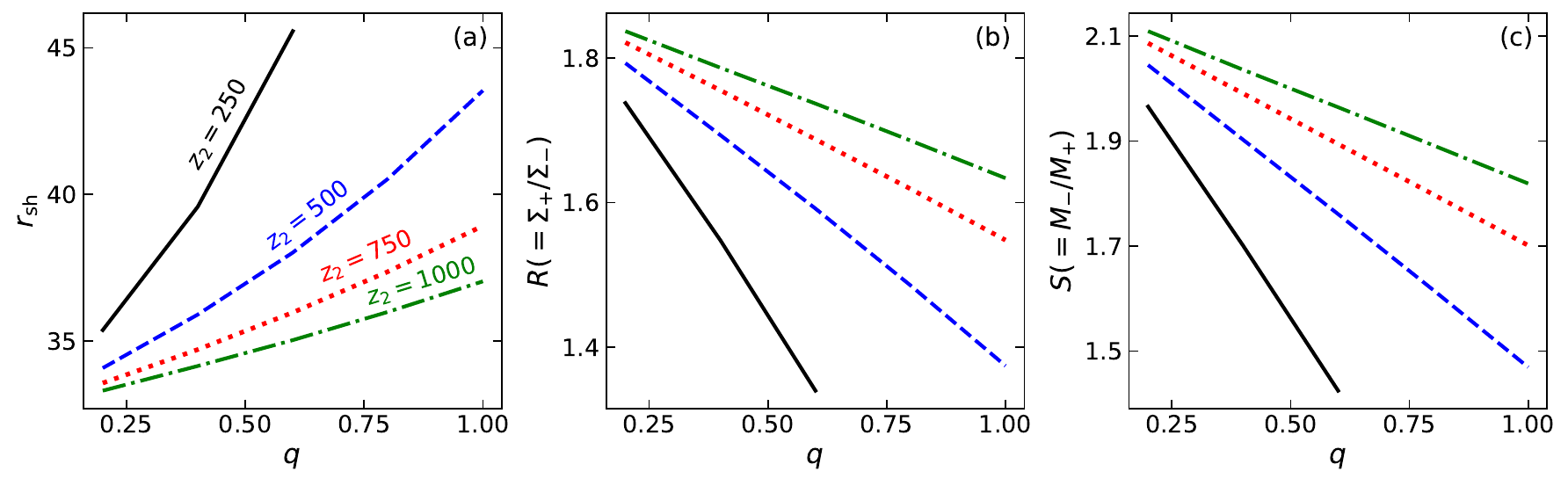}
	\caption{Variation of shock location ($r_{\text{sh}}$) in panel (a), compression ratio ($R$) in panel (b), and shock strength ($S$) in panel (c) as a function of the mass ration ($q$) for different binary separations $z_2 = 250$, $500$, $750$, and $1000$. In this figure, we set $\lambda = 3.5$ and $E = 1.01$. See the text for details.}
	\label{fig:sp}
\end{figure*}

Next, we examine the shock properties in terms of the binary parameters $q$ and $z_2$. To do this, we calculate the shock radius ($r_{\text{sh}}$) as a function of $q$ for different values of $z_2$. The obtained results are presented in Fig. \ref{fig:sp}a, where the solid (black), dashed (blue), dotted (red), and dash-dotted (green) curves correspond to $z_2 = 250$, $500$, $750$, and $1000$, respectively. In this case, the input parameters are chosen as $\lambda = 3.5$ and $E = 1.01$. We observe that the shock radius recedes away from the black hole as $q$ increases, irrespective of the value of $z_2$. Moreover, the shock disappears when $q$ exceeds a critical value, as the shock conditions are no longer satisfied beyond that limit. Also, for a given $q$, the shock radius decreases when $z_2$ increases. Since both the density and temperature of the flow undergo significant changes at the shock location, it is essential to study their variations across the shock locations. To examine the density compression, we define a quantity, which is called the compression ratio as $R = \Sigma_{+}/\Sigma_{-}$, where $``+"$ and $``-"$ denote the post-shock and pre-shock values of the surface density $\Sigma$ ($= 2\rho H$). Following Eq. (\ref{eq:mass_accretion}), we find an expression of $R$ in terms of flow radial 3-velocity ($v$) and its Lorentz factor ($\gamma_{v}$) as $R = (v\gamma_{v})_{-}/(v\gamma_{v})_{+}$. In Fig. \ref{fig:sp}b, we present the variation of $R$ with $q$ for the same set of input parameters used in Fig. \ref{fig:sp}a. It is observed that $R$ decreases with $q$ for a given value of $z_2$, which is expected, as the shock locations shift to larger radii when $q$ increases. Since the Mach number ($M$) depends on the flow temperature ($\Theta$), the temperature jump at the shock front is characterized by a quantity called the shock strength ($S$). It is defined as the ratio of the pre-shock Mach number ($M_{-}$) to the post-shock Mach number ($M_{+}$) as $S = M_{-}/M_{+}$. In Fig. \ref{fig:sp}c, we show the variation of $S$ with $q$. As the shock radius increases with $q$, it is again expected that $S$ decreases as $q$ increases. Furthermore, for a fixed $q$, both $R$ and $S$ increase with $z_2$ as $r_{\text{sh}}$ shrinks with $ z_2$. It is noted that the dependencies of the shock location, density compression, and temperature compression on both $q$ and $z_2$ are consistent with the results presented in Figs. \ref{fig:sh-solutions-diff-q} and \ref{fig:sh-solutions-diff-z2}.  

\subsection{$E - \lambda$ parameter space for shock solutions}
\label{sec:parameter-space}

\begin{figure*}
	\centering
	\includegraphics[width=\linewidth]{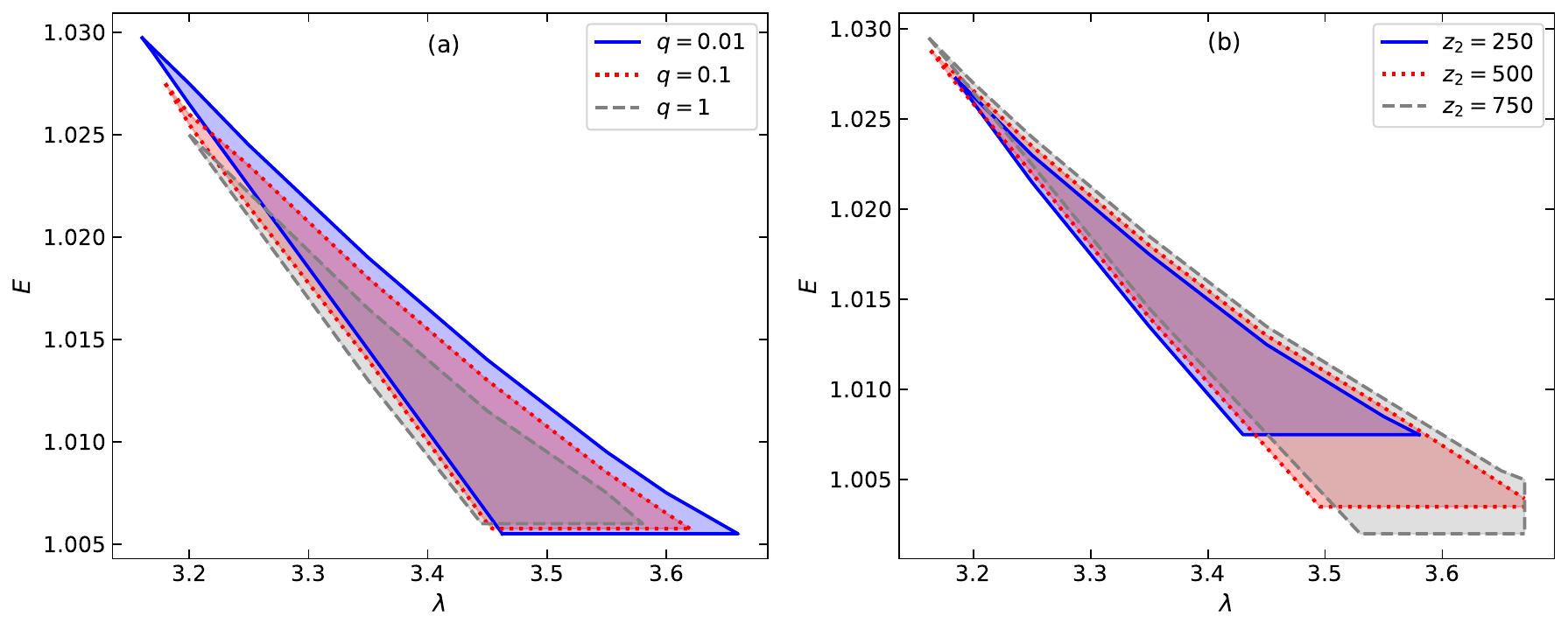}
	\caption{Modification of the shock parameter space in the specific angular momentum ($\lambda$) and energy ($E$) plane for the mass ratios $q = 0.01$, $0.1$, and $1$ with binary separation $z_2 = 300$ (panel (a)), and for $z_2 = 250$, $500$, and $750$ with $q = 0.5$ (panel (b)). See the text for details.}
	\label{fig:ps-shock}
\end{figure*}

We mention earlier that the A-type solutions can harbor shock transitions, provided the relativistic shock conditions are satisfied. Most importantly, shock solutions are not unique but rather have an energy range for a given angular momentum (see Section \ref{sec:critical-points-analysis}). Here, we explore the available parameter space for shock solutions in the CPD and see the modification of those identified regions with the mass ratio ($q$) and binary separation ($z_2$). For this, we calculate an effective region of the parameter space in the specific angular momentum ($\lambda$) and energy ($E$) plane for a given combination of $q$ and $z_2$ that admits shock solutions. In Fig. \ref{fig:ps-shock}a, we present the modification of the shock parameter space for different values of $q$ with a fixed $z_2 = 300$. The regions bounded by the solid (blue), dotted (red), and dashed (gray) curves correspond to the results for $q = 0.01$, $0.1$, and $1$, respectively. We observed that, the parameter space shifts toward lower $\lambda$ and lower $E$ domains as $q$ increases. Moreover, the area under the parameter space gradually decreases when the mass of the secondary black hole becomes comparable to that of the primary black hole. Similarly, in Fig. \ref{fig:ps-shock}b, we show the modification of the shock parameter space in the $\lambda-E$ plane as a function of $z_2$ with a fixed $q = 0.5$, where the effective regions bounded by the solid (blue), dotted (red), and dashed (gray) are obtained for $z_2 = 250$, $500$, and $750$, respectively. It is observed that as $z_2$ increases, the parameter space shifts toward higher $\lambda$ and lower $E$ sides. In addition, the shock parameter space expands as the secondary black hole moves farther away from the primary one.

\section{Conclusions}
\label{sec:conclusions}

In this work, we study the accretion flow properties of the circumprimary disc (CPD) in a binary black hole system. We derive the model equations for general relativistic accretion flow in the background of a binary black hole spacetime. These equations are solved numerically to obtain the transonic accretion solutions both in presence and absence of shocks. Since the secondary black hole can influence the horizon area of the primary black hole, the main objective of this work is to examine the impact of the secondary black hole, mainly its mass and distance with respect to the primary black hole, on the accretion properties of the CPD.

We find that flow may possess single or multiple critical points depending on the specific angular momentum ($\lambda$), energy ($E$), binary mass ratio ($q$), and binary separation ($z_2$). We obtain the A-type accretion solutions that contain multiple critical points and show that the behaviors of the accretion solution solutions remain A-type with an increase of both $q$ and $z_2$. Since the A-type accretion solutions can exhibit standing shock transitions, provided the relativistic shock conditions are satisfied, we also investigate the shock-induced accretion solutions. We analyze the shock properties such as shock radius ($r_{\text{sh}}$), compression ratio ($R$), and shock strength ($S$) as functions of $q$ and $z_2$. We observe that as $q$ increases, both $R$ and $S$ decreases as $r_{\text{sh}}$ moves away from the horizon. However, as $z_2$ increases, $r_{\text{sh}}$ decreases, resulting in higher values of $R$ and $S$. We further investigate the luminosity distribution of the CPD as a function of $q$ and $z_2$. We find that the CPD spectra barely depend on $q$ and $z_2$ for both the shock-induced and shock-free solutions, which agrees with the findings in \cite{Lee-2024-141}. Also, we examine the shock parameter space in the $\lambda - E$ plane and observe how it is modified by $q$ and $z_2$. We find that as $q$ increases, the shock parameter space shrinks, whereas it expands with an increase in $z_2$. Therefore, the possibility of shock transitions is greater for smaller mass ratios and larger binary separations. Indeed, this analysis implies that while the shock properties of the CPD moderately change due to the presence of a secondary black hole, it does not significantly affect the CPD spectrum.

It is worth noting that a recent study by \cite{Lee-2024-141} investigated the spectral features of BBH accretion flows. The authors analyzed CBD accretion and its associated spectrum in the presence of irradiation from CPD and CSD, treating all disc components as standard discs within a Newtonian framework. Additionally, the accretion discs were assumed to be coplanar. In their study, the authors showed that the CBD spectrum is nearly independent of the binary mass ratio. However, in our study, we focus solely on the accretion properties of the CPD within a fully general relativistic framework. Since the present symmetry of the BBH metric (Eq. (\ref{eq:BBH-metric})) allows the fluid to rotate around the Z-axis to conserve angular momentum, the accretion discs of the two black holes would not be coplanar in this scenario. Additionally, we consider the CPD as an optically thin medium that does not emit like a black body, as is typically assumed in standard disc models. Therefore, although our analysis, including the disc configuration, differs entirely from that of \cite{Lee-2024-141}, the observations regarding the binary mass ratio are qualitatively similar.

Finally, we mention the limitations of our work. We do not incorporate viscosity \cite[]{Dihingia-2019-2412, Patra-2024-371}, magnetic fields \cite[]{Mitra-2022-5092, Mitra-2024-28, Dihingia-2024-4}, thermal conduction \cite[]{Mitra-2023-4431, Singh-2024-02256}, and radiative cooling \cite[]{Sarkar-2018-1950037, Sarkar-2022-34} in the flow equations. Moreover, we only consider the thermal bremsstrahlung emission, neglecting the synchrotron and Compton emission processes \cite[]{Dihingia-2020-3043, Sarkar-2022-34}. However, their presence in an accretion model is expected to provide more valuable insights into accretion physics. Also, we consider a binary system consisting of two Schwarzschild black holes. In reality, astrophysical black holes possess some spin. Exact analytical solutions for a stable binary configuration of two spinning black holes are already available in the literature \cite[]{Astorino-2022-829}. We expect, the spin parameters may also effect the CPD accretion flow. We plan to investigate all these aspects in future work and report our findings elsewhere. 

\section*{Data availability statement}
The data underlying this article will be available with reasonable request.

\section*{Acknowledgments}
The authors thank the anonymous reviewers for their valuable comments and useful suggestions, which have improved the quality of the paper. The work of SP is supported by the University Grants Commission (UGC), India, under the National Eligibility Test (NET) - Senior Research Fellowship (SRF) scheme. BRM is supported by a START-UP RESEARCH GRANT from the Indian Institute of Technology Guwahati, India, under the grant SG/PHY/P/BRM/01. SD is supported by the Science and Engineering Research Board (SERB), India, through the grant MTR/2020/000331.        

\bibliographystyle{apsrev}
\bibliography{references} 

\end{document}